\documentclass{article}
\usepackage{graphicx}
\usepackage[round]{natbib}
\usepackage{epsfig}
\title{
An objective change-point analysis of historical Atlantic hurricane numbers
}

\author{
Stephen Jewson (RMS)\footnote{\emph{Correspondence email}: \texttt{stephen.jewson@rms.com}}\\
Jeremy Penzer (LSE)\\}

\begin{document}
\maketitle

\begin{abstract}
We perform an objective change-point analysis on 106 years of historical
hurricane number data. The algorithm we use looks at all possible combinations of
change-points and compares them in terms
of the variances of the differences between real and modelled numbers.
Overfitting is avoided by using cross-validation.
We identify four change-points, and show that the presence of temporal
structure in the hurricane number time series is highly statistically significant.
\end{abstract}

\section{Introduction}

Several severe hurricanes made landfall on the US coastline during
2004 and 2005, and this has increased the level of interest in
questions related to long-term fluctuations in the levels of
hurricane activity. One way to contribute to an overall
understanding of hurricane activity levels is to analyse the
historical hurricane record statistically, and over the years this
has been attempted by a number of authors (for example,
see~\citet{goldenberg}, \citet{elsnerj00} and \citet{elsnern04}).
The main questions that are typically considered are: is the
record stationary, and if not, what does the variability look
like? A number of studies have concluded that the record is
not stationary, and that there are periods of high and low levels
of activity, although the precise causes of these fluctuations are
not agreed upon in detail.


Exactly when the periods of high and low activity start and end,
and how to identify these start and end points, is also not
exactly clear.
In \citet{elsnerj00}, a change-point scheme
based on log-linear regression was used to examine the major
Atlantic hurricane time series. In \citet{elsnern04} a change-point analysis based on a Markov Chain Monte Carlo approach
\citep{lavielle} is used to analyse both basin-wide Atlantic
hurricane activity and US landfalling rates. While our primary
interest in this article is Atlantic activity, we note that a
number of studies have analysed various Pacific tropical-cyclone time series using
log-linear regression and Bayesian techniques \citep{chu02, chu04,
zhao}.

In this paper we revisit the question of how to detect change-points in Atlantic hurricane activity: our contribution is to use
what we consider to be better statistical methods for the
identification of periods of high and low activity than have been
used before. We think that the methods we use are more or less the
best that one can hope to do: we look at all possible combinations
of different positions for changes in the level of activity, and
compare the resulting models using cross-validation to avoid
overfitting. These methods are now possible because of recent
increases in available computer power.

\section{Methods}

\begin{figure}[!hb]
  \begin{center}
    \scalebox{1.2}{\includegraphics{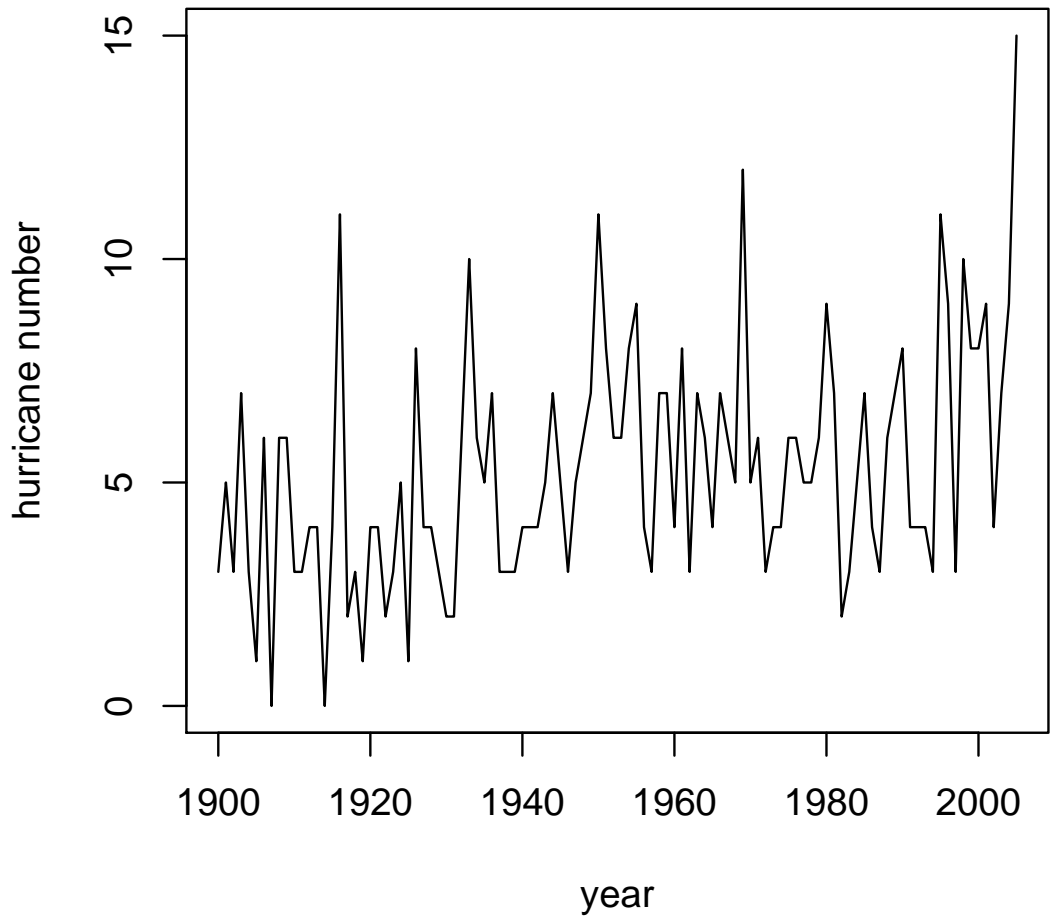}}
  \end{center}
    \caption{
Atlantic basin hurricane numbers for the period 1900 to 2005. }
     \label{f01}
\end{figure}

Our method for identifying different levels of activity in the
historical hurricane data works as follows. For data, we take the
numbers of Atlantic hurricanes per year as reported in the current
version of the HURDAT database~\citep{hurdat}. This data runs from
1869 to 2005, although we only consider data from 1900 to 2005
because of doubts about the completeness of the data prior to
1900. The data from 1900 to 2005 is shown in figure~\ref{f01}. One
might also have doubts about the data for the period 1900-1949,
prior to the use of aerial reconnaissance: however, we analyse the
data as-is. All of our conclusions must be considered with this in
mind.

We model this series of hurricane numbers using sequences of
levels of constant hurricane activity, plus noise. Initially we
model the series as a simple constant level, with no
change-points, then using two constant levels, with a single
change-point. For all possible positions of this single change-point we calculate the predictive mean square error (MSE) and we
consider the best model to be that which minimises this MSE score.
We say \emph{predictive} mean square error because we calculate
the MSE using cross-validation, thus avoiding overfitting. Not
using cross-validation would unfairly favour the selection of
small gaps between change-points.

We then increase the number of possible change-points and repeat
this exercise. As the number of change-points is increased one
might expect the MSE results to improve, as we model the real
fluctuations in the series, but then at some point one might
expect the MSE results to get worse, as the model becomes
overfitted. The model we choose is the one with lowest MSE.

The only parameter in the model is the minimum gap between change-points.
We start by trying a gap of 2 years, and then increase the gap to 10 years,
for reasons discussed below.

\section{Results: 2 year minimum gap}

\begin{figure}[!hb]
  \begin{center}
    \scalebox{1.2}{\includegraphics{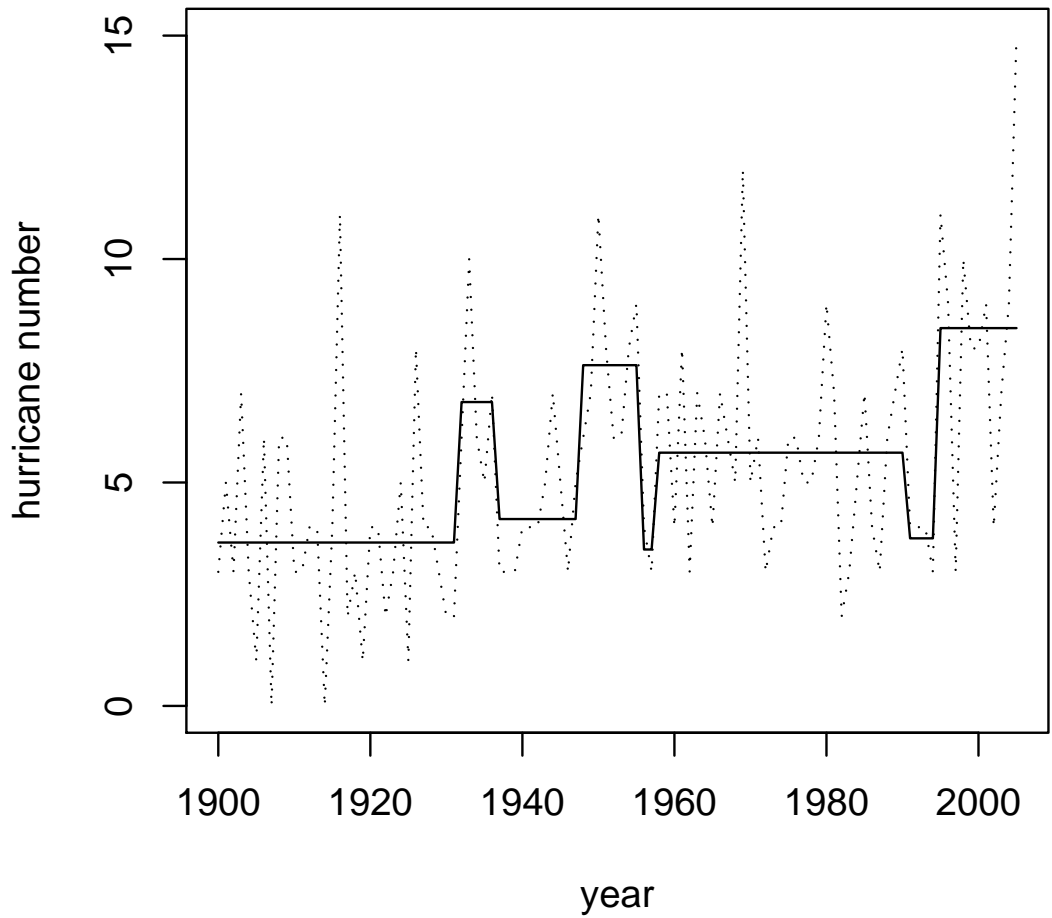}}
  \end{center}
    \caption{
Change-points for the 8 level model with minimum gap of 2 years.}
     \label{2f01}
\end{figure}

We first consider results from our change-point analysis
for when the minimum gap between change-points is set to two years.
Table~\ref{2t01} shows the change-points identified in this case, for models with numbers of change-points increasing
from 0 (1 level) to 7 (8 levels). `48' in this table indicates that a change-point has been identified
between 1947 and 1948.
Table~\ref{2t02} shows the predictive RMSE scores for these
models, and table~\ref{2t03} shows the number of combinations of
change-points tested in each case. The predictive RMSE scores
decrease as the number of levels increases, right up to the last
case tested, which has 8 levels and 7 change-points. From
table~\ref{2t03} we see that testing 8 levels requires
consideration of over 50 trillion combinations, and this reaches
the limit of our computing power. We cannot claim, therefore, that
we have been able to find the best model, since there might be a
better model with 9 levels (or even more than 9 levels). The change-points
detected in the 8 level case show an interesting distribution in
time, with small gaps between several of the pairs (in spite of
the fact that we are using cross-validation). The change-points
identified in the 8 level model are depicted in figure~\ref{2f01},
against the hurricane time series data. Overall the results
suggest that the hurricane number time series is not stationary,
and that the underlying rate undergoes fluctuations on a range of
timescales.

At this point we are forced to conclude that our 2-year minimum
gap analysis has failed because we have been unable to identify a
global minimum in our cost function for lack of computer power.
For many purposes, however, we are less interested in identifying
very short time-scale fluctuations in hurricane rates than we are
in understanding longer time-scale fluctuations. For this reason,
we now increase the minimum gap allowed from 2 years to 10 years,
in order to focus on fluctuations on decadal and longer
timescales.

\section{Results: 10 year minimum gap}

\begin{figure}[!hb]
  \begin{center}
    \scalebox{0.8}{\includegraphics{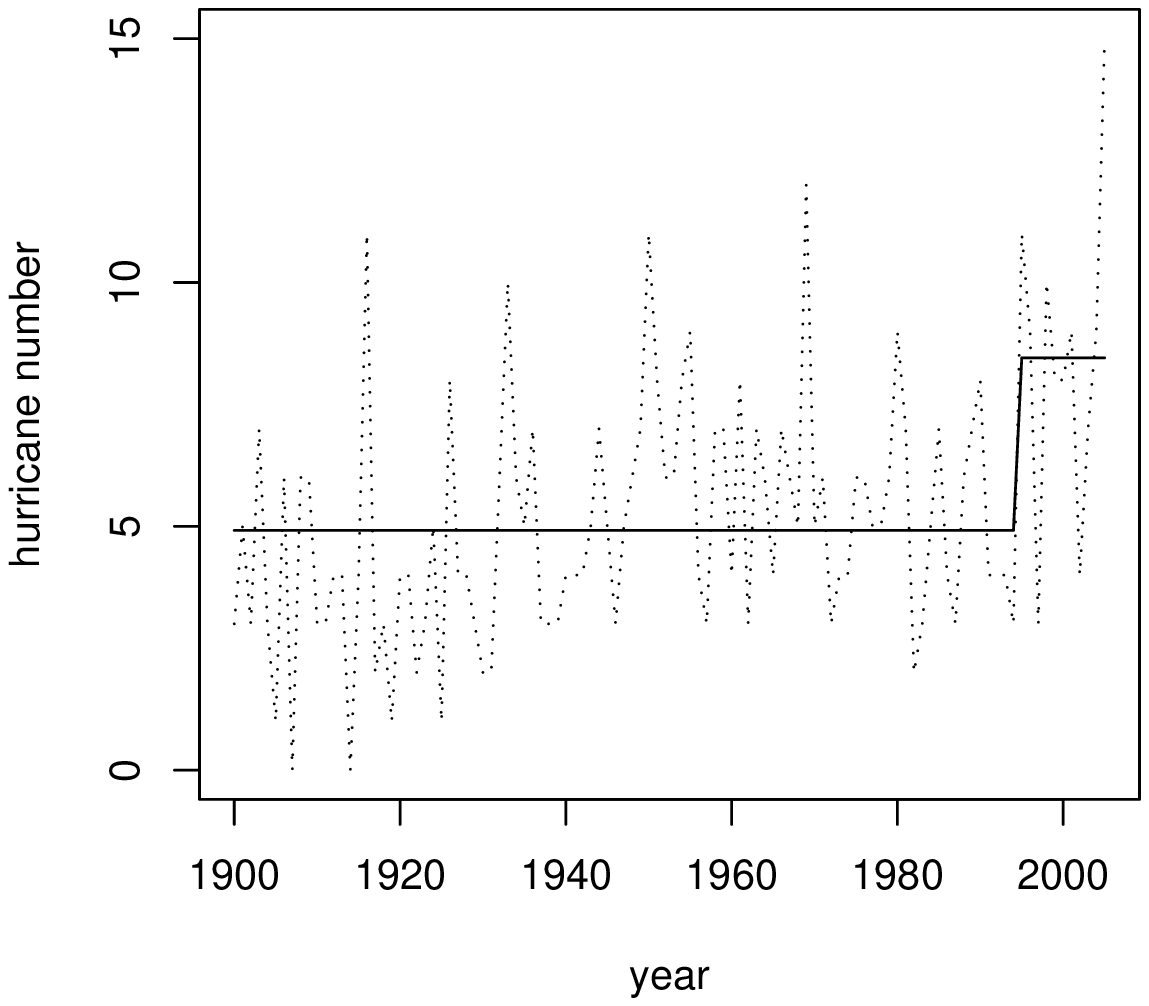}}
  \end{center}
    \caption{
The best 2 level model (for a 10 year minimum gap). }
     \label{10f01}
  \begin{center}
    \scalebox{0.8}{\includegraphics{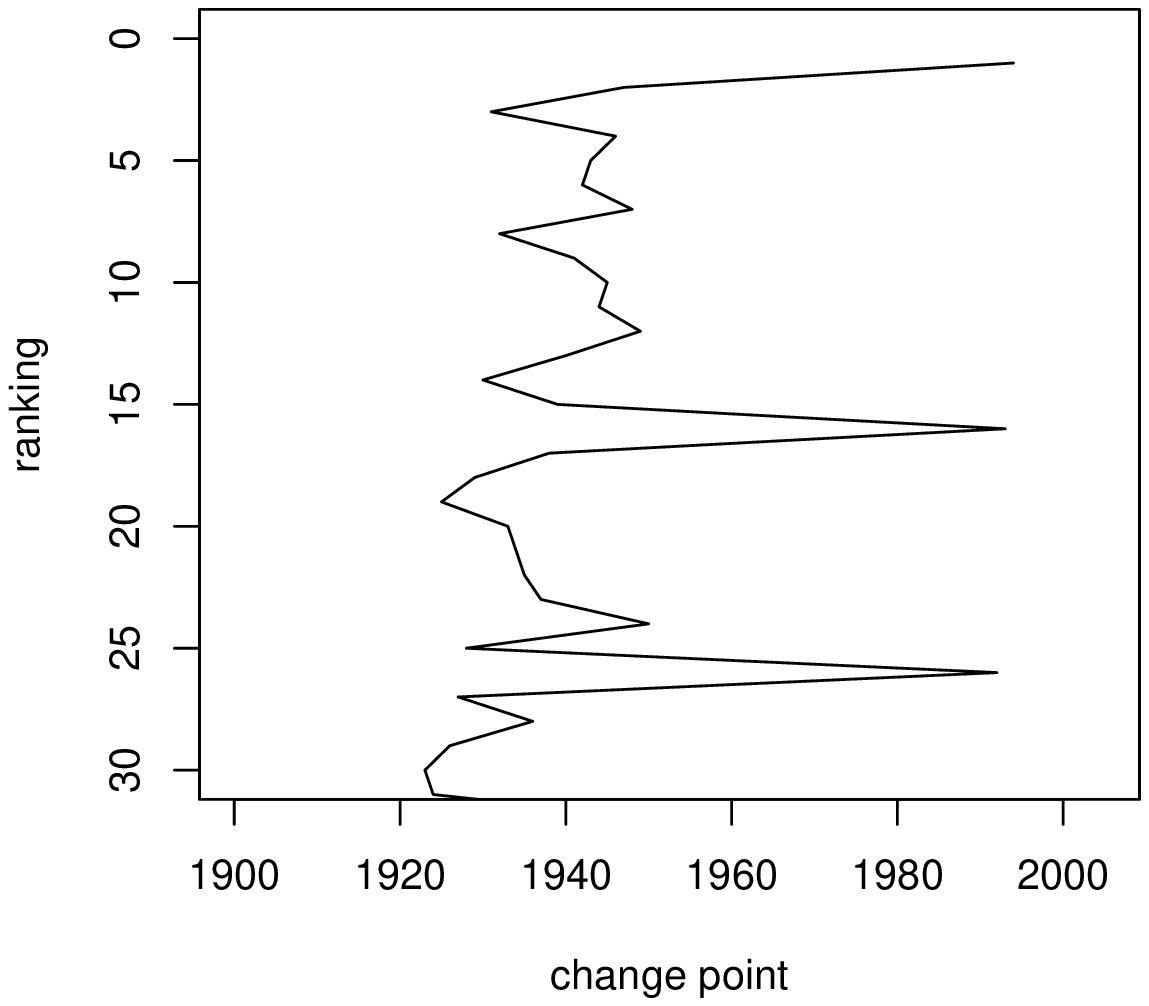}}
  \end{center}
    \caption{
The change-points for the top 30 two level models considered. }
     \label{10f02}
\end{figure}

We now consider results from our change-point analysis for when the minimum gap between
change-points is set to ten years. Tables~\ref{10t01}, \ref{10t02}
and~\ref{10t03} show the change-points, scores and numbers of
combinations considered in this case. Looking at the scores, we
now see that we reach a minimum RMSE score for 5 levels and 4
change-points. For a greater number of levels the RMSE increases,
indicating that the model then starts to become overfitted
relative to the 4 change-point model. We illustrate the change-points for the cases with 2 to 5 levels, from figure~\ref{10f01}
onwards. In each case we also show the change-points for all of
the top 30 combinations identified, which gives some idea of the
robustness of the results. Interestingly for the 5 level case the
change-points seem to be very robust, and very similar sets of
change-points occur several times in the top 30 results.

\section{Significance testing}

Could these results have occurred if the data were purely random?
We test this as follows. We take the historical hurricane number data used in our change-point analysis, and create
100 random reorderings. Each of these reordered time-series has
the same marginal distribution as the original data, but different
temporal structure. We then apply our change-point algorithm to
each of these 100 series. The results are as follows. With respect
to the number of change-points we identify (by the first minimum
in the series of RMSE values): on average, we find 4.5 change-points, with a range from 2 to 7. This tells us that the fact that
we have identified 4 change-points in the real series is not
itself an indication of real temporal structure. With respect to
the RMSE values achieved: the average of the 100 minimum RMSE
values achieved is 2.60, while the \emph{lowest} of the 100 values
is $2.41$. This is \emph{larger} than the value achieved from the
real data, which is 2.31. This shows, with a high level of
certainty, that the RMSE score result for 4 change-points from the real data could
not have occurred from random data, and is very strong evidence
that there is real temporal structure in the hurricane number
time-series. We note, however, that we have not proven that the
change-points we have identified are definitely right, or even
statistically significant, on an individual basis. Many of the
individual combinations of change-points
we have tested are statistically significant, but the
differences between them are not. All we can say for sure is:
\begin{itemize}
    \item we have proven that there is decadal time-scale variability
    in the time-series
    \item that the best way to approximate that variability, within the class
    of models we have considered, is
    given by the change-points that we have detected
    \item if one has to choose one set of change-points, the change-points we
    have detected are probably the best set to choose
\end{itemize}

\section{Intense hurricanes}

\begin{figure}[!hb]
  \begin{center}
    \scalebox{1.2}{\includegraphics{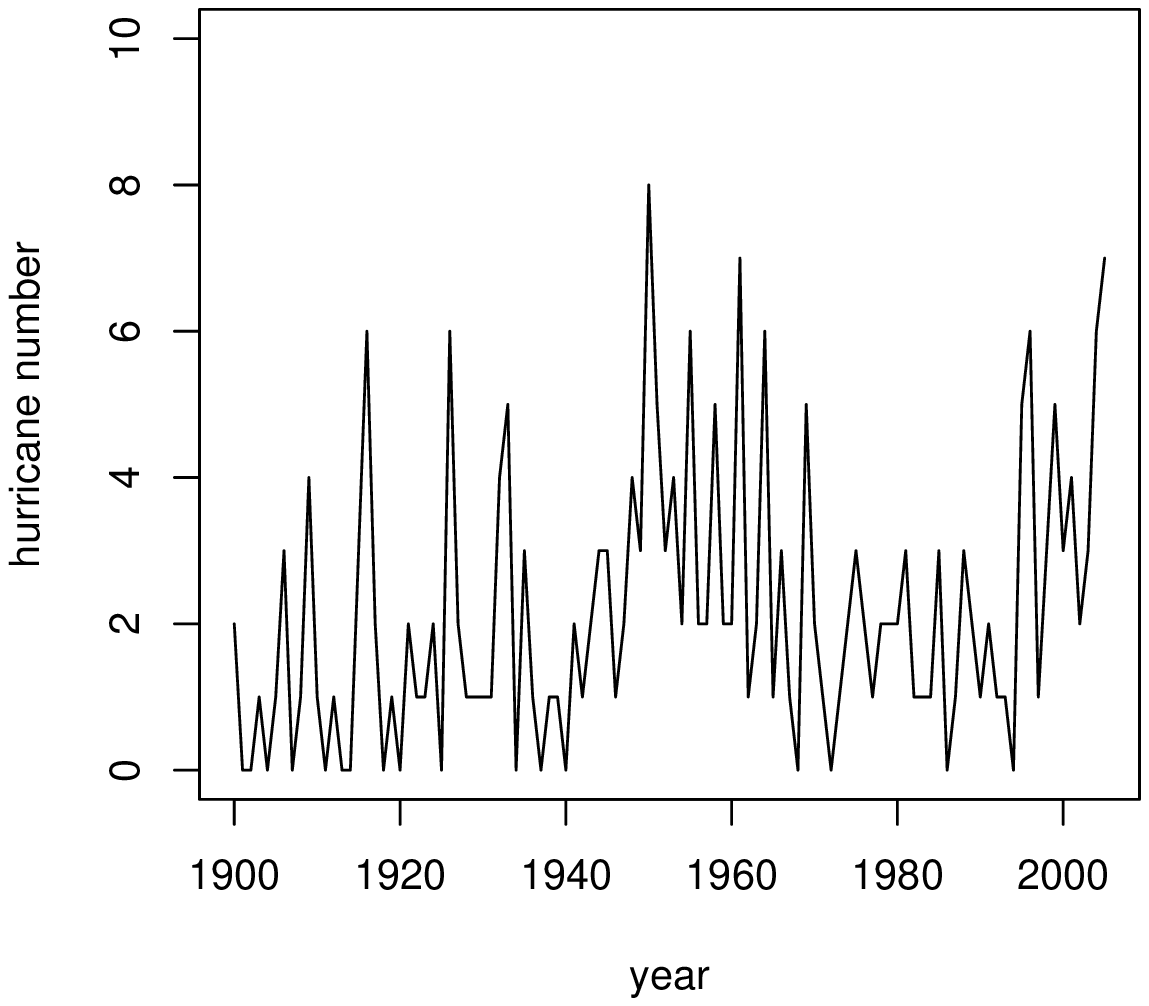}}
  \end{center}
    \caption{
Atlantic basin \emph{intense} hurricane numbers for the period
1900 to 2005. }
     \label{if01}
\end{figure}

Up to now our analysis has focussed on the identification of
change-points in the time-series of the total number of
hurricanes. However, \citet{elsnerj00} and \citet{elsnern04}
consider only the intense hurricanes (Category 3-5 on the
Saffir-Simpson scale).
It is therefore of interest to run our new
algorithm on the intense hurricanes only, to understand whether
the differences between our results and those of Elsner are mainly a result of
using a different data set (all storms versus cat 3-5 storms) or because we use a different
algorithm. Figure~\ref{if01} shows the time-series of intense hurricane
numbers. By eye, the change-points look more significant than
those in the time-series of all hurricane numbers shown in
figure~\ref{f01}.
Tables~\ref{it01} and \ref{it02} show the change-points and scores
for our analysis of the intense hurricane number time-series, with
a minimum window width of 10 years, as before. We see that the lowest score
is once again at 4 change-points (5 levels). Relative to the change-points we
identified in the basin series, two are exactly the same (48 and
95), one has `moved' a little (70 to 65), and one has changed (32
to 15). We note that for the same time series,  \citet{elsnerj00}
and \citet{elsnern04} identify 3 change-points, at 43, 65 and 95.
Our analysis identifies 2 identical change-points (65 and 95), and 1 that is
close (43 for Elsner et al. versus 48 in our analysis). Our analysis has revealed an additional
change-point at 1915, presumably because we are using a different search algorithm,
although because of uncertainty in the earlier data one must have significant
doubts as to whether this change-point has any physical significance.
By and large, our analyses for major hurricanes is very similar to
previous studies.

Our conclusions from the comparison between our results for cat 1-5 and cat 3-5
hurricanes and the results in~\citet{elsnerj00} for cat 3-5 hurricanes are that
(a) the change-point in 1994/1995
is robust to changing between cat 1-5 and cat 3-5 , and to changing detection
methods,
(b) the 1964/1965-1969/1970 change-point occurs in 1969/1970 for
cat 1-5 data and in 1964/1965 for cat 3-5 data. In the cat 3-5 data it is robust
to the use of different detection methods,
(c) the change-points earlier in the century are not robust
to the use of different detection methods.

\section{Discussion}

We have completed a new change-point analysis of the hurricane number time series from 1900 to 2005.
We consider the method that we have used to be close to being the best that one could possibly do,
since we consider all possible combinations of change-points. The method also has the advantage that
it is conceptually very simple. The only disadvantage is that a vast number of computations are required.
The one parameter in the model is the minimum gap allowed between change-points. Setting
this to 2 years makes the problem computationally unfeasible for us, since we don't find an optimum
solution before the number of combinations becomes too large to search in a reasonable time on our computer.
Increasing the parameter to 10 years, and thus focussing on fluctuations on time-scales of decades
and longer, reduces the number of combinations and turns out to be computationally tractable.
We find that the absolute global optimum solution to this problem has 4 change-points and 5 different levels.
The change-points occur at 1931/1932, 1947/1948, 1969/1970 and 1994/1995.

When we reapply the method to intense hurricanes only we again
find an absolute global optimum solution with 4 change-points and
5 different levels. Two of the change-points are the same as for the
total hurricane number series (1947/1948 and 1994/1995), one has
moved a little earlier (1969/1970 becomes 1964/1965) and one has
changed (1931/1932 becomes 1914/1915). This final change-point
should be viewed with a lot of suspicion, however, since the data
is considered rather unreliable this early in the century.
The two most recent change-points we have found in the intense
time-series agree exactly with the two most recent change-points
found in earlier work (using a different algorithm) by~\citet{elsnerj00}.
This is perhaps not
that surprising: in the intense time-series, at least, one can more or less
identify the change-points by eye.

This study is our first attempt at looking at change-points in the hurricane number time series. There
are various directions in which we plan to take this research, such as considering a probabilistic
cost function, applying the same analysis to landfalling hurricane numbers, and using the results
to predict future levels of hurricane activity.

\section{Acknowledgements}

Thanks to Manuel Lonfat, Roman Binter and Shree Khare for interesting discussions on this topic, and thanks
to Alexandra Guerrero for helping us run the computer code.

\bibliography{arxiv}

\newpage

\newpage
\begin{table}[h!]
  \centering
\begin{tabular}{|c|c|c|c|c|c|c|c|}
 \hline
 1 & 2 & 3 & 4 & 5 & 6 & 7 &8\\
 \hline
 model& cp1 & cp2 & cp3 & cp4 & cp5 & cp6 & cp7\\
 \hline
1&&&&&&&\\
2&  48&&&&&&\\
3&  32&  95&&&&&\\
4&  48&  56&  95&&&&\\
5&  32&  49&  56&  95&&&\\
6&  32&  37&  48&  56&  95&&\\
7&  32&  37&  48&  56&  91&  95&\\
8&  32&  37&  48&  56&  58&  91&  95\\

 \hline
\end{tabular}
\caption{ The change-points identified in the hurricane number
time-series, versus the number of levels, for a minimum gap of 2
years. }\label{2t01}
\end{table}
\begin{table}[h!]
  \centering
\begin{tabular}{|c|c|}
 \hline
 1 & 2\\
 \hline
 model& predictive RMSE\\
 \hline
 1&    2.673664     \\
 2&    2.467585     \\
 3&    2.349803     \\
 4&    2.317592     \\
 5&    2.289907     \\
 6&    2.254462     \\
 7&    2.231958     \\
 8&    2.216252     \\

 \hline
\end{tabular}
\caption{ The predictive RMSE scores for the different
models.}\label{2t02}
\end{table}
\begin{table}[h!]
  \centering
\begin{tabular}{|c|c|}
 \hline
 1 & 2\\
 \hline
 model& number of combinations tested\\
 \hline
1&  1.00E+00\\
2&  1.03E+02\\
3&  1.02E+04\\
4&  9.70E+05\\
5&  8.85E+07\\
6&  7.74E+09\\
7&  6.47E+11\\
8&  5.28E+13\\

 \hline
\end{tabular}
\caption{ The number of combinations tested for each
model.}\label{2t03}
\end{table}

\newpage

\newpage
\begin{table}[h!]
  \centering
\begin{tabular}{|c|c|c|c|c|c|c|}
 \hline
 1 & 2 & 3 & 4 & 5 & 6 & 7\\
 \hline
 model& cp1 & cp2 & cp3 & cp4 & cp5 & cp6\\
 \hline
1&&&&&&\\
2&  48&&&&&\\
3&  32&  95&&&&\\
4&  32&  82&  95&&&\\
5&  32&  48&  70&  95&&\\
6&  32&  48&  70&  82&  95&\\
7&  17&  32&  48&  70&  82&  95\\

 \hline
\end{tabular}
\caption{ The change-points identified in the hurricane number
time-series, versus the number of model levels, now for a longer
minimum gap of 10 years.}\label{10t01}
\end{table}

\begin{table}[h!]
  \centering
\begin{tabular}{|c|c|}
 \hline
 1 & 2\\
 \hline
 model& predictive RMSE\\
 \hline
 1&   2.673664    \\
 2&   2.467586    \\
 3&   2.349803    \\
 4&   2.335162    \\
 5&   2.314494    \\
 6&   2.316886    \\
 7&   2.326645    \\

 \hline
\end{tabular}
\caption{ The predictive RMSE scores for the different
models.}\label{10t02}
\end{table}
\begin{table}[h!]
  \centering
\begin{tabular}{|c|c|}
 \hline
 1 & 2\\
 \hline
 model& number of combinations tested\\
 \hline
1&  1.00E+00\\
2&  8.70E+01\\
3&  5.93E+03\\
4&  3.01E+05\\
5&  1.06E+07\\
6&  2.29E+08\\
7&  2.57E+09\\

 \hline
\end{tabular}
\caption{ The number of combinations tested for each
model.}\label{10t03}
\end{table}

\newpage

\newpage
\begin{figure}[!hb]
  \begin{center}
    \scalebox{0.8}{\includegraphics{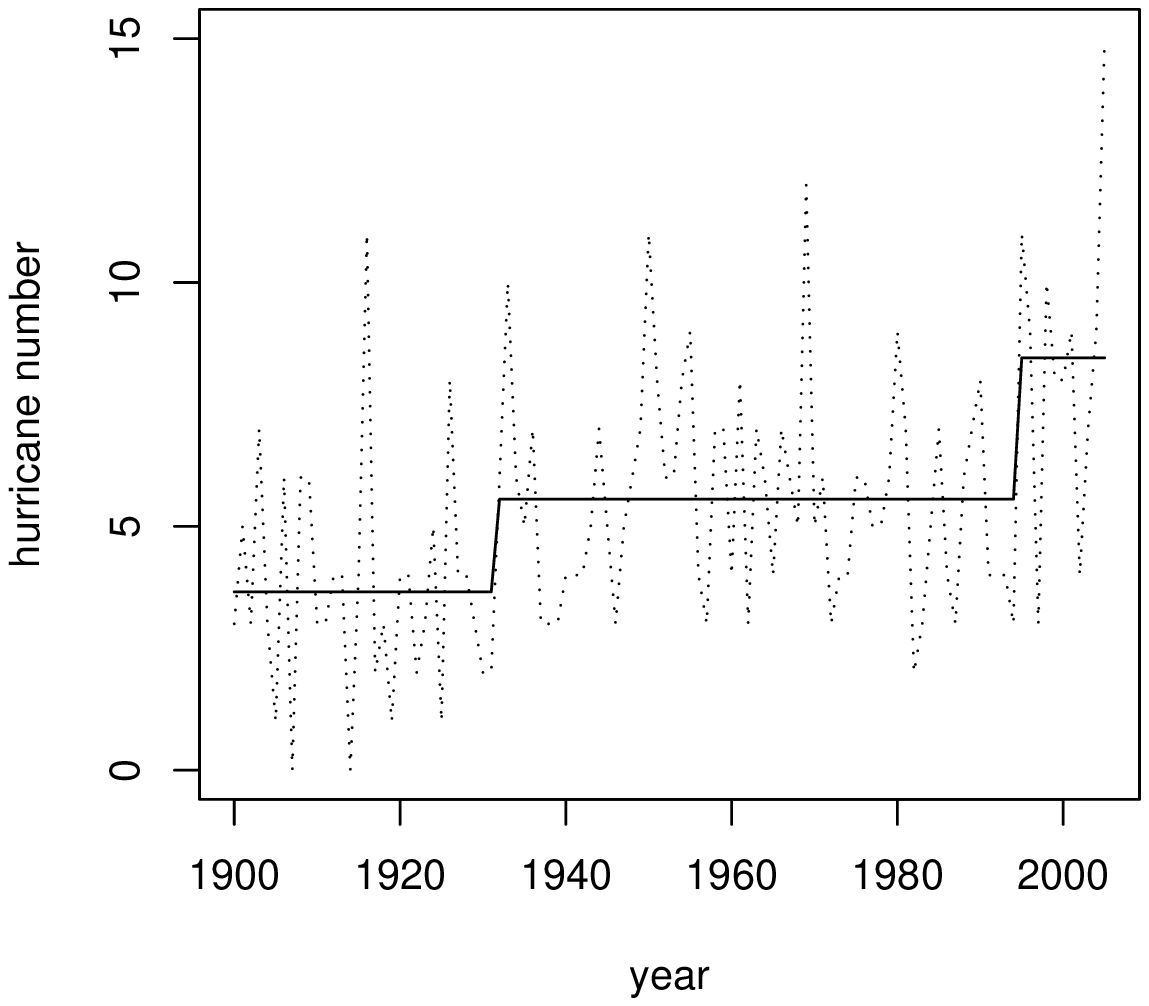}}
  \end{center}
    \caption{
The best 3 level model (for a 10 year minimum gap). }
     \label{10f03}
  \begin{center}
    \scalebox{0.8}{\includegraphics{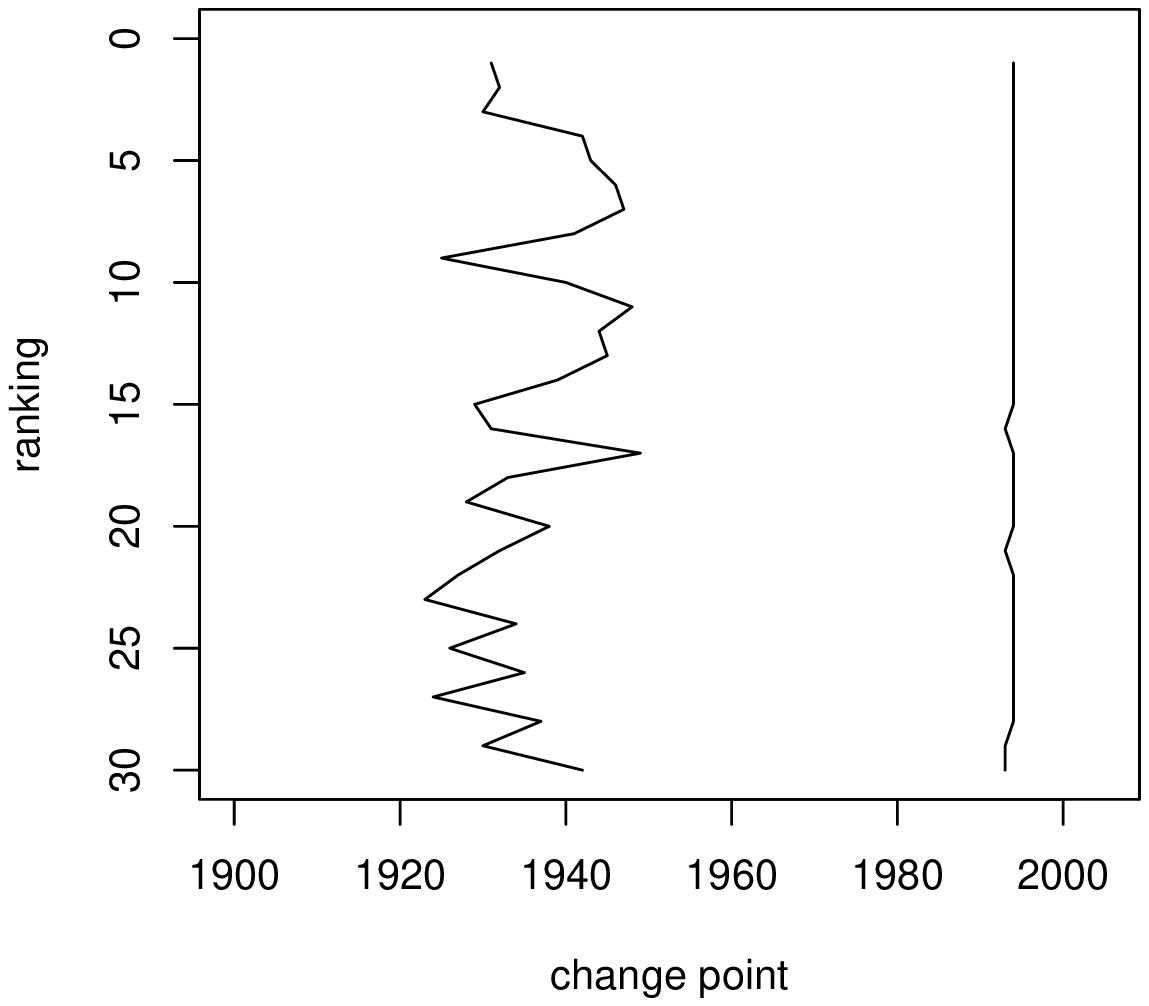}}
  \end{center}
    \caption{
The change-points for the top 30 three level models. }
     \label{10f04}
\end{figure}

\newpage
\begin{figure}[!hb]
  \begin{center}
    \scalebox{0.8}{\includegraphics{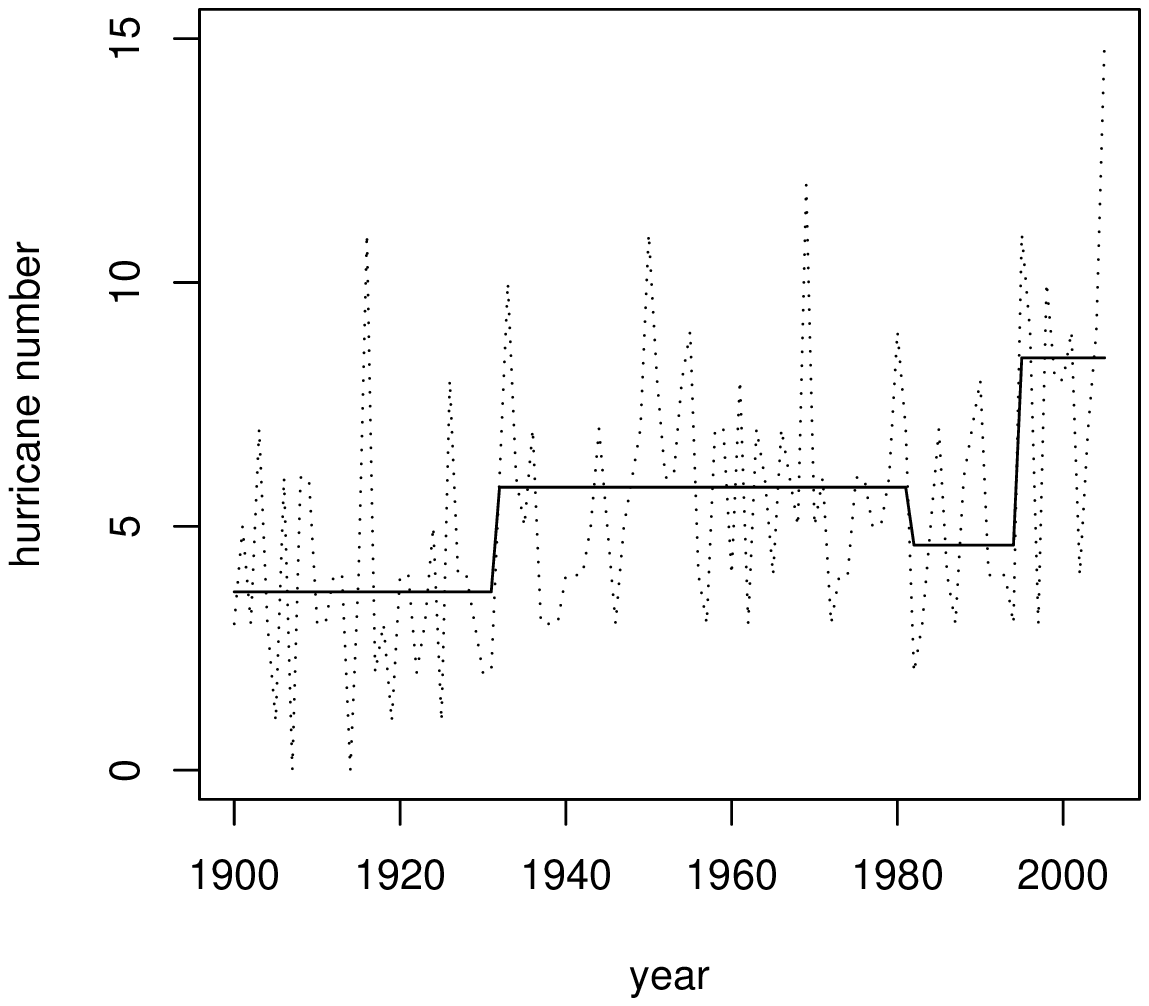}}
  \end{center}
    \caption{
The best 4 level model (for a 10 year minimum gap). }
     \label{10f05}
  \begin{center}
    \scalebox{0.8}{\includegraphics{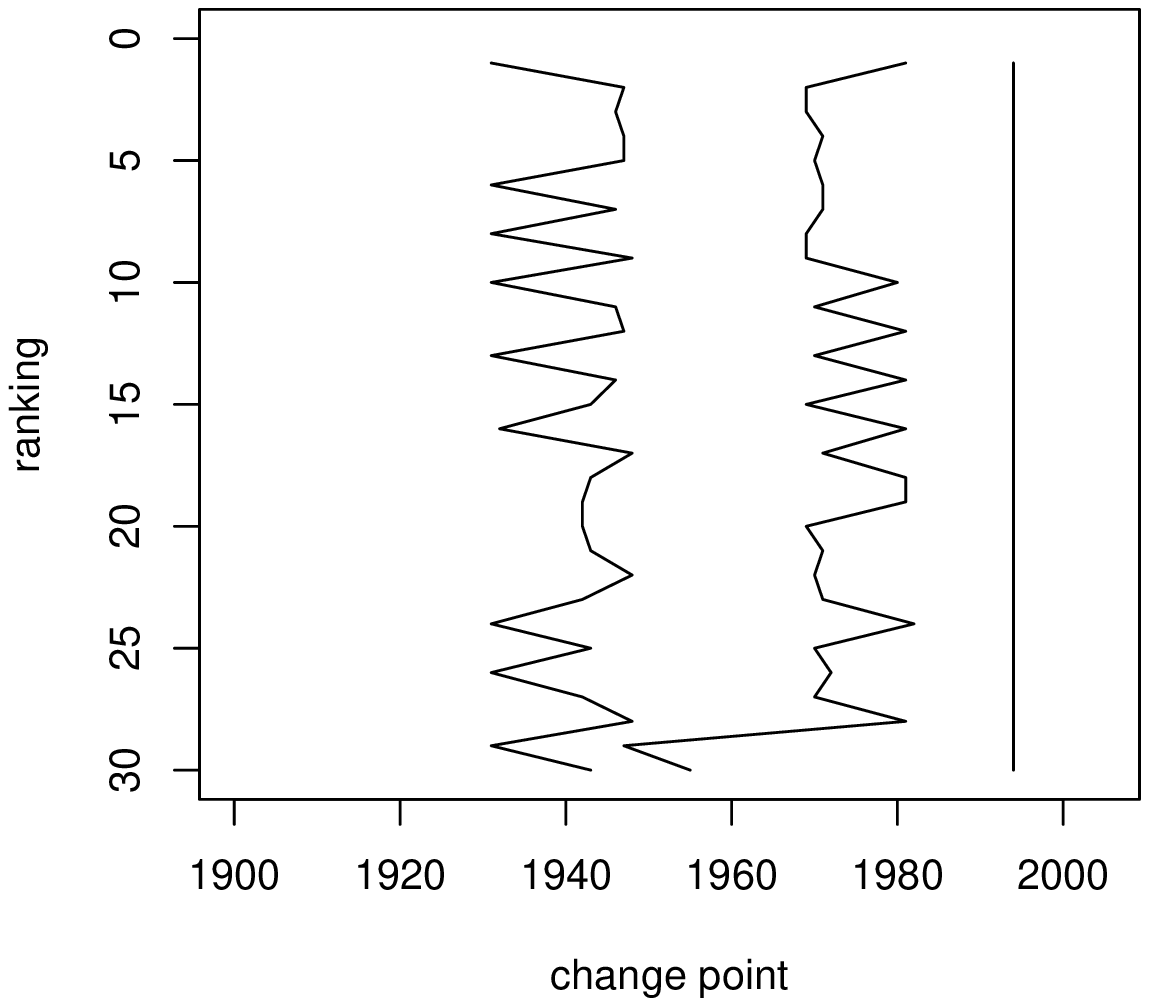}}
  \end{center}
    \caption{
The change-points for the top 30 four level models. }
     \label{10f06}
\end{figure}

\newpage
\begin{figure}[!hb]
  \begin{center}
    \scalebox{0.8}{\includegraphics{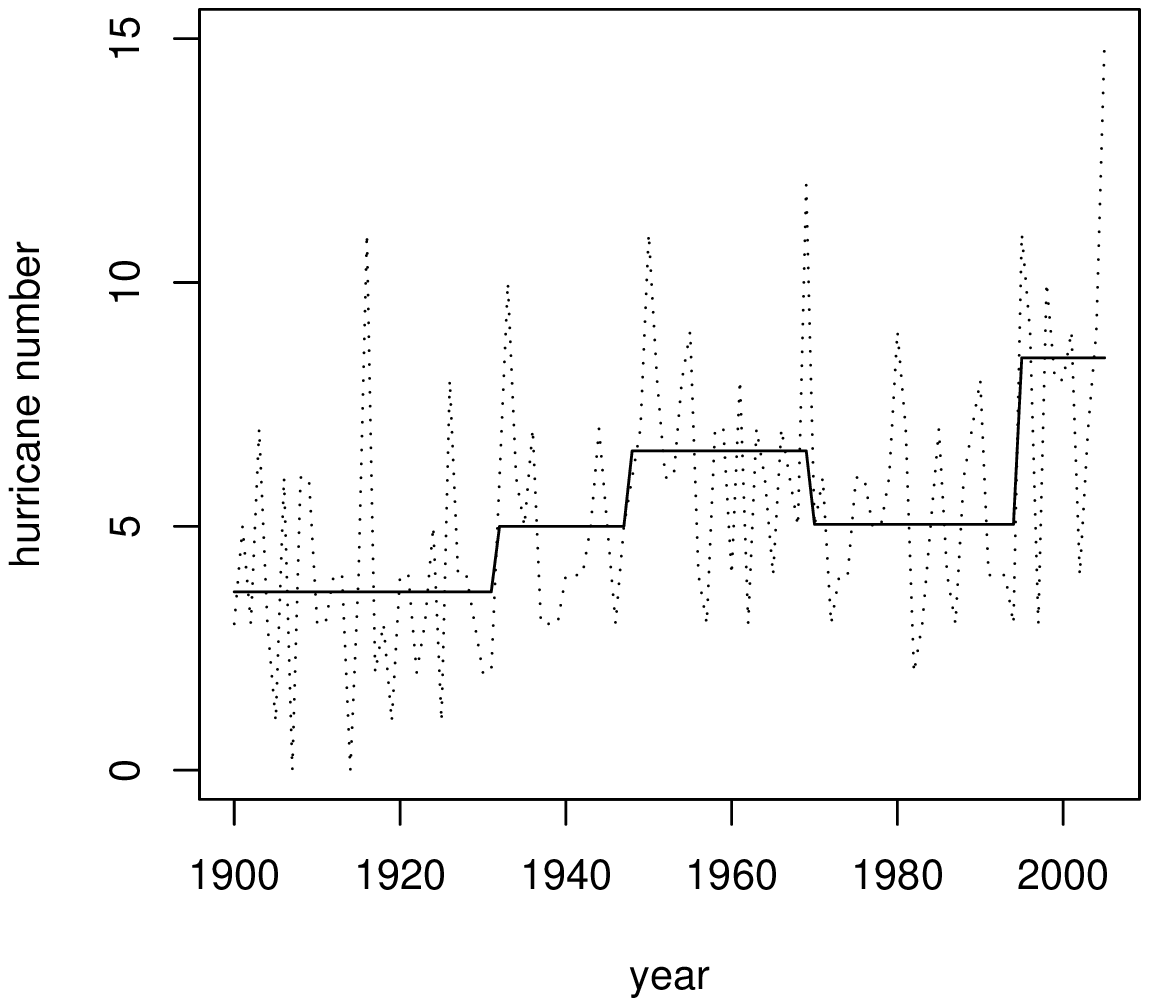}}
  \end{center}
    \caption{
The best 5 level model (for a 10 year minimum gap). }
     \label{10f07}
  \begin{center}
    \scalebox{0.8}{\includegraphics{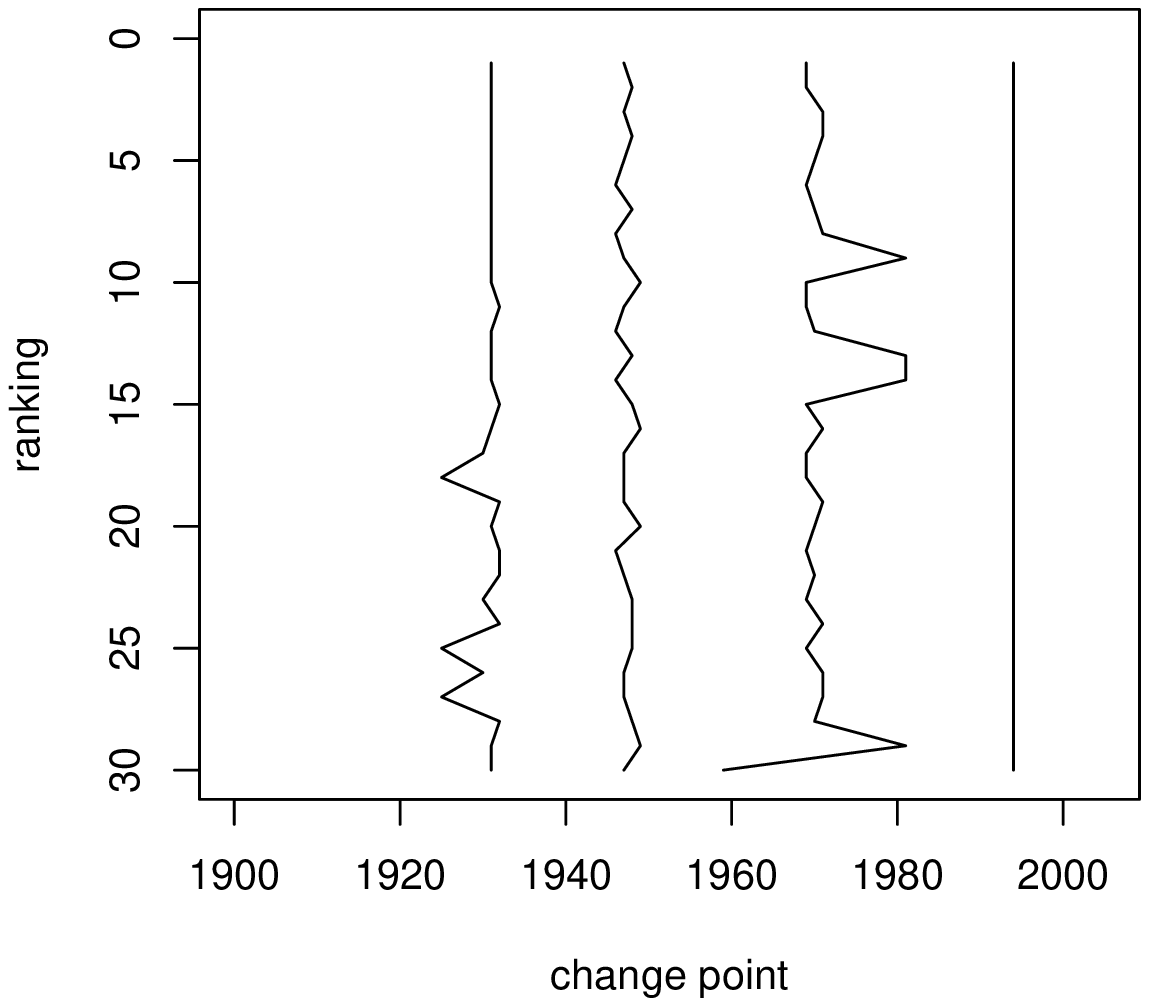}}
  \end{center}
    \caption{
The change-points for the top 30 five level models. }
     \label{10f08}
\end{figure}

\newpage

\newpage
\begin{table}[h!]
  \centering
\begin{tabular}{|c|c|c|c|c|c|c|}
 \hline
 1 & 2 & 3 & 4 & 5 & 6 & 7\\
 \hline
 model& cp1 & cp2 & cp3 & cp4 & cp5 & cp6\\
 \hline
1&&&&&&\\
2&  95&&&&&\\
3&  15&  95&&&&\\
4&  48&  65&  95&&&\\
5&  15&  48&  65&  95&&\\
6&  15&  48&  65&  82&  95&\\
7&  15&  36&  48&  65&  82&  95\\

 \hline
\end{tabular}
\caption{ The change-points identified in the \emph{intense}
hurricane number time-series, versus the number of levels, for a
minimum gap of 10 years. }\label{it01}
\end{table}
\begin{table}[h!]
  \centering
\begin{tabular}{|c|c|}
 \hline
 1 & 2 \\
 \hline
 model& predictive RMSE\\
 \hline
 1&   1.878292    \\
 2&   1.778038    \\
 3&   1.737661    \\
 4&   1.615974    \\
 5&   1.604409    \\
 6&   1.605744    \\
 7&   1.609706    \\

 \hline
\end{tabular}
\caption{ The predictive RMSE scores for the different
models.}\label{it02}
\end{table}

\newpage
\begin{figure}[!hb]
  \begin{center}
    \scalebox{0.8}{\includegraphics{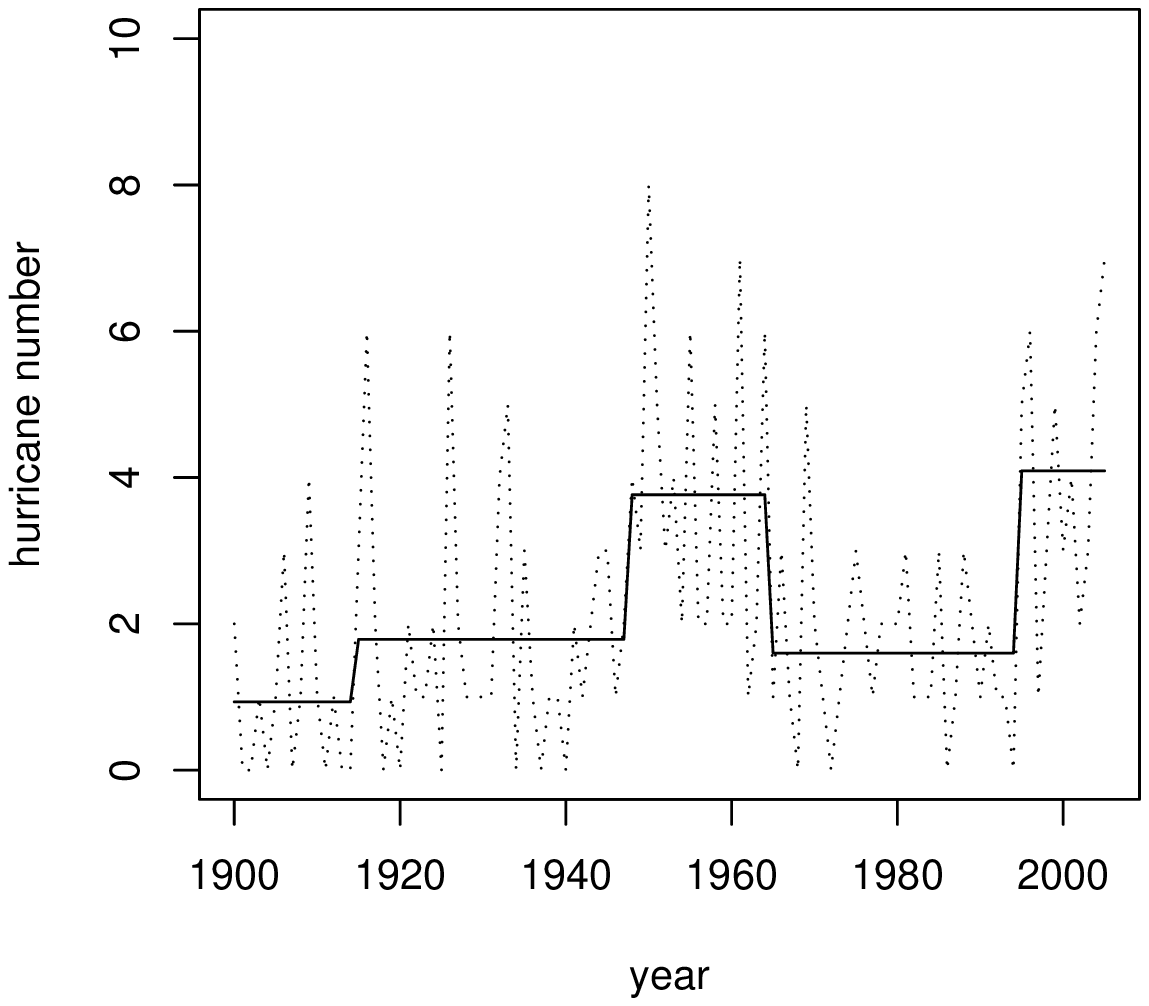}}
  \end{center}
    \caption{
The best 5 level model (for a 10 year minimum gap). }
     \label{if07}
  \begin{center}
    \scalebox{0.8}{\includegraphics{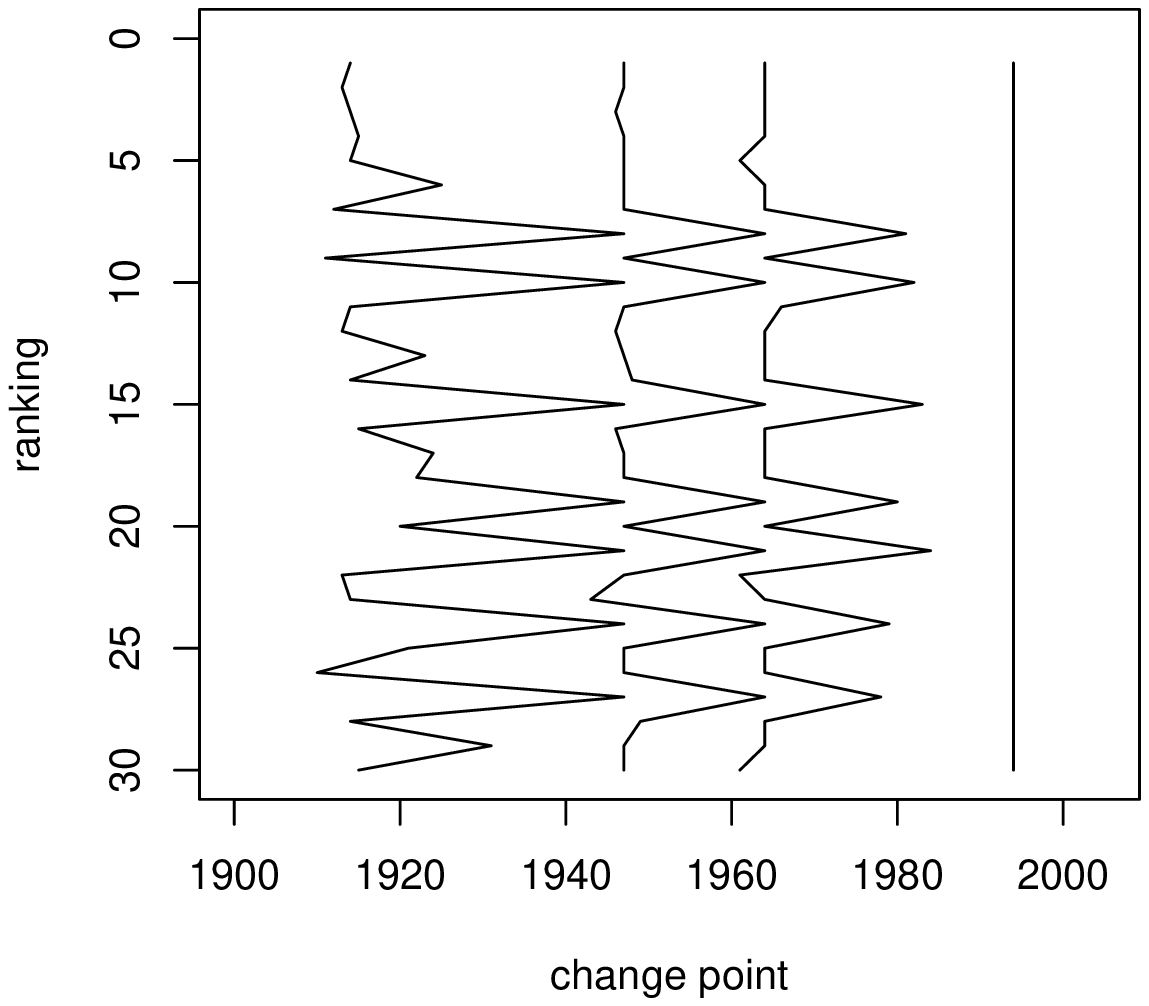}}
  \end{center}
    \caption{
The change-points for the top 30 five level models. }
     \label{if08}
\end{figure}

\end{document}